\documentclass[aps,prl,twocolumn,showpacs,superscriptaddress]{revtex4}  
\usepackage{graphicx}    
\usepackage{dcolumn}     
\usepackage{bm}          
\usepackage{amssymb}     
\begin{document}


\title{Luttinger liquid behavior in weakly disordered quantum wires}
\author{E.~Levy}%
\email{eyallevy@post.tau.ac.il} \affiliation{School of Physics and
Astronomy, Raymond and Beverly Sackler Faculty of Exact Sciences,
Tel Aviv University, Tel Aviv 69978, Israel}
\author{A.~Tsukernik}%
\affiliation{University Research Institute for Nanoscience and
Nanotechnology, Tel Aviv University, Tel Aviv 69978, Israel}
\author{M.~Karpovski}%
\affiliation{School of Physics and Astronomy, Raymond and Beverly
Sackler Faculty of Exact Sciences, Tel Aviv University, Tel Aviv
69978, Israel}
\author{A.~Palevski}%
\affiliation{School of Physics and Astronomy, Raymond and Beverly
Sackler Faculty of Exact Sciences, Tel Aviv University, Tel Aviv
69978, Israel}
\author{B.~Dwir}%
\affiliation{Ecole Polytechnique Federale de Lausanne (EPFL),
Laboratory of Physics of Nanostructures, CH-1015 Lausanne,
Switzerland}
\author{E.~Pelucchi}%
\affiliation{Ecole Polytechnique Federale de Lausanne (EPFL),
Laboratory of Physics of Nanostructures, CH-1015 Lausanne,
Switzerland}
\author{A.~Rudra}%
\affiliation{Ecole Polytechnique Federale de Lausanne (EPFL),
Laboratory of Physics of Nanostructures, CH-1015 Lausanne,
Switzerland}
\author{E.~Kapon}%
\affiliation{Ecole Polytechnique Federale de Lausanne (EPFL),
Laboratory of Physics of Nanostructures, CH-1015 Lausanne,
Switzerland}
\author{Y.~Oreg}%
\affiliation{Department of Condensed Matter Physics, The Weizmann
Institute of Scienece, Rehovot 76100, Israel}

\begin{abstract}
We have measured the temperature dependence of the conductance in
long V-groove quantum wires (QWRs) fabricated in GaAs/AlGaAs
heterostructures. Our data is consistent with recent theories
developed within the framework of the Luttinger liquid model, in the
limit of weakly disordered wires. We show that for the relatively
low level of disorder in our QWRs, the value of the interaction
parameter $g\cong0.66$, which is the expected value for GaAs.
However, samples with a higher level of disorder show conductance
with stronger temperature dependence, which does not allow their
treatment in the framework of perturbation theory. Fitting such data
with perturbation-theory models leads inevitably to wrong (lower)
values of $g$.
\end{abstract}

 \maketitle


The electrical conductance through noninteracting clean quantum
wires (QWRs) containing a number of one-dimensional subbands is
quantized in the universal unit $\frac{2e^{2}}{h}$ \cite{Landauer},
as observed in narrow constrictions in 2D electron gas (2DEG)
systems \cite{Wees,Wharam}. For such short and clean narrow wires,
the e-e interactions described by the so-called Luttinger liquid
(LL) model \cite{Tomonaga} do not affect the value of the
conductance, namely it is temperature and length independent as
indeed was shown experimentally \cite{Wees,Wharam}. In the presence
of disorder in sufficiently long QWRs, suppression of the
conductance is expected at low temperatures. A number of theoretical
papers addressing this issue \cite{Kane,Fukuyama,Maslov1,Oreg}
predict a negative correction to the conductance versus temperature
$G(T)$, which increases with $T$ and obeys a power law: $T^{g-1}$,
where $g<1$ is an interaction parameter.


The validity of the implications of the LL theory has been recently
demonstrated in a number of experiments \cite{Yacoby1,McEuen}. The
most evident proofs of the predictions were shown in
\emph{tunnelling} experiments performed in T-shaped cleaved-edged
overgrown GaAs quantum wires \cite{Yacoby1} and in carbon nanotubes
\cite{McEuen}. Earlier \emph{non-tunnelling} experiments, in which
suppression of conductance occurs in the linear response regime, did
not unambiguously prove the validity of the theory, and the value of
the $g$ parameter could not be deduced from the experimental data
\cite{Tarucha,Yacoby2,Rother}. Several complications are encountered
in such experiments. For sufficiently disordered wires, where the
correction to $G(T)$ is expected to be large, the value of the
conductance at the plateau is not well defined due to the specific
realization of the disordered potential in the wire, as was the case
for the long wires of Tarucha \emph{et al.} \cite{Tarucha}.
Moreover, in the intermediate regime, namely for disorder level for
which the conductance plateau could be well defined but the
corrections to $G(T)$ are already significant for a relatively
narrow temperature range, $g$ cannot be extracted by applying a
perturbation theory. If however, the disorder is very weak so that
the plateaus are well defined at all temperatures
\cite{Tarucha,Yacoby2}, the variation of its value versus
temperature is so weak that the $g$ parameter cannot be reliably
determined. Therefore, if one wishes to compare $G(T)$ to the
theory, a wire possessing just the right amount of disorder is
needed.

In this work, we present an experimental study of the conductance in
single mode V-groove GaAs QWRs. The variation of conductance was
measured over a wide temperature range. Our results are consistent
with the theories \cite{Maslov1,Oreg} based on the LL model for
weakly disordered wires, allowing us to deduce the value of
$g=0.66$, as expected for interacting electrons in GaAs and as was
observed experimentally in tunneling experiments \cite{Yacoby1}. We
show results for QWRs displaying different amounts of disorder, thus
enabling us to show the importance of the degree of disorder and the
limits of perturbation theory.

\begin{figure}[t]
\centering
\includegraphics[height=2.3in,width=1\linewidth]{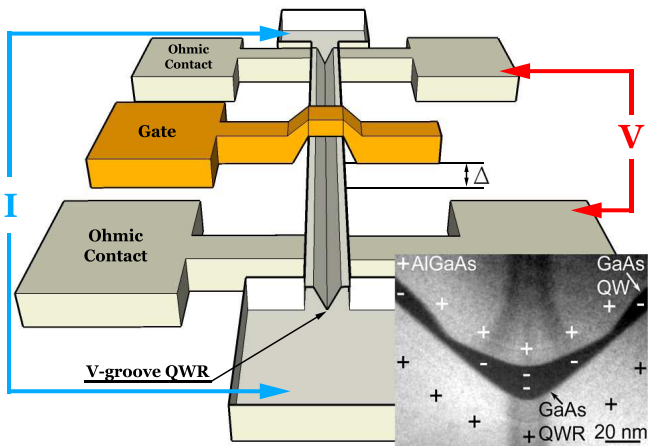}
\caption{Schematic diagram of the device's geometry. The QWR is
located at the bottom of the V-groove. The inset shows  a
cross-sectional TEM image of the wire, on which the charge
distribution is schematically depicted.}\label{sample}
\end{figure}

The QWRs studied here were produced by low pressure ($20~mbar$)
metalorganic vapor phase epitaxy (MOVPE) of GaAs/AlGaAs
heterostructures on undoped (001) GaAs substrates patterned with
V-grooves oriented in the [01-1] direction, fabricated by
lithography and wet chemical etching \cite{Kapon}. The
heterostructure consisted of a $230~nm$ GaAs buffer layer, $1.1~\mu
m$ $Al_{0.27}Ga_{0.73}As$ lower barrier layer, $14~nm$ GaAs quantum
well (QW) layer, $160~nm$ $Al_{0.27}Ga_{0.73}As$ upper barrier
layer, and a $10~nm$ GaAs cap layer. All layers were nominally
undoped, except for two $20~nm$ Si doped
$\left(\approx1\times10^{18}~cm^{-3}\right)$ regions in the
$Al_{0.27}Ga_{0.73}As$ barriers, spaced by $80$ and $60~nm$,
respectively, from the lower and upper GaAs QW interface, serving as
modulation doping regions. The layer thicknesses refer to growth on
a planar (100) sample. Growth of the GaAs QW in V-grooves yields a
crescent-shaped QWR flanked on both sides by \{111\}A oriented QWs
(see inset of Fig. \ref{sample}). The modulation doping yields a 1D
electron gas confined to the wire, laterally connected to 2DEG
systems that form on the \{111\}A QWs.

The QWRs were contacted using the scheme illustrated schematically
in Fig. 1. Source and drain Au/Ge/Ni pads were fabricated using
standard photolithography techniques with a mesa etched along the
QWR, providing ohmic contacting to the 2DEG regions. Additionally,
narrow ($0.5~\mu m$) Ti/Au Schottky gates were formed using electron
beam lithography in order to isolate the QWR and control the number
of populated 1D subbands in it.

The conductance was measured by the four-terminal method using a low
noise analog lock-in amplifier (EG\&G PR-124A). The excitation
current was kept at $I=0.1~nA$ ensuring that the voltage drop across
the wire never exceeded $\frac{k_{B}T}{e}$ at the lowest
temperature. Without application of gate voltage $V_{g}$, the
transport in our system is carried by electrons in the 2DEG on the
sidewalls and in parallel with those in the 1D QWR. Application of a
sufficiently large negative $V_{g}$ depletes the electrons at the
sidewalls and creates a 1DEG confined to the V-groove QWR underneath
the gate \cite{Kaufman1}. At a certain range of still more negative
voltage, a single populated 1D channel is realized. As was
demonstrated \cite{Kaufman2}, the electrons remain at their
one-dimensional state during a transition length $\Delta$ (see Fig.
\ref{sample}) on both sides of the gate. This transition length
arises from the weak coupling between the 1D states and the located
2DEG, which acts as an electron reservoir. This transition length,
defined as the length required for electrons to be scattered
into/from the 2DEG, was found to be as large as $\Delta=2~\mu m$
\cite{Kaufman2}. It is thus reasonable to conclude that the
effective length of the 1D wire exceeds the actual width of the gate
($0.5~\mu m$)by about $2~\mu m$ on each side of the gate.
\begin{figure}[b]
\centering
\includegraphics[width=1\linewidth]{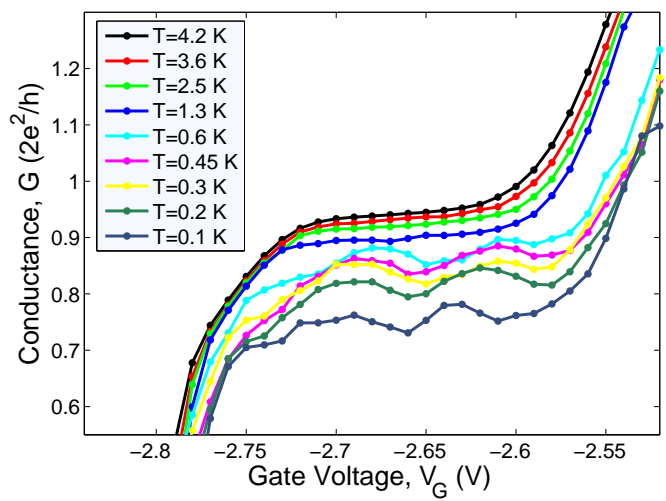}
\caption{Conductance vs. gate voltage $V_{g}$ for $0.5~\mu m$ gate
width at various temperatures, after subtraction of series
resistance.}\label{Plateau}
\end{figure}

\begin{figure}[b]
\centering
\includegraphics[width=1\linewidth]{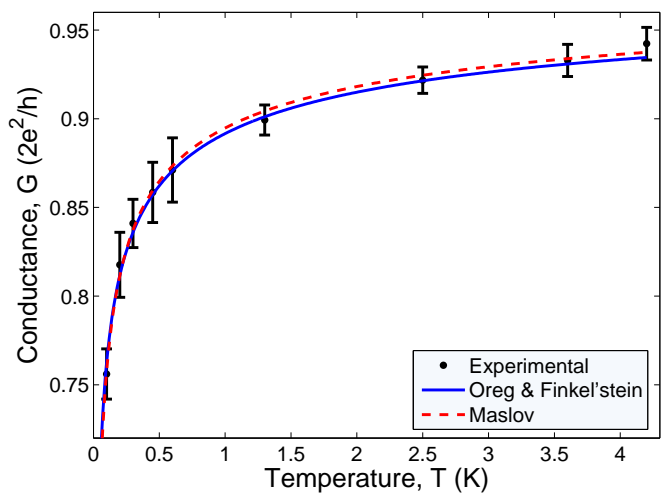}
\caption{Conductance values of the first plateau vs. temperature in
the wire of Fig. \ref{Plateau} (points with error bars). Both
theoretical expressions are plotted for the same parameters, e.g.
$g=0.64$ and $T_{0}=0.7~mK$ of equation (\ref{Oreg}).}\label{Inset}
\end{figure}

Fig. \ref{Plateau} shows the variation of the conductance with gate
voltage $V_{g}$, in the range where electrons populate only a single
1D sub-band, at temperatures between $100~mK$ and $4.2~K$. The
temperature was measured using a calibrated carbon thermometer
(Matsushita 56 resistor). The electronic temperature of GaAs 2DEG
does not deviate from the bath temperature for $T>100~mK$ as shown
in our previous studies \cite{Shkol}. The data was taken at
stabilized temperatures of the bath while the $V_{g}$ was swept
through the entire range. A series resistance of $180~\Omega$,
measured at $V_{g}=0$, has been subtracted from all curves. At
$4.2~K$ the conductance plateau is smooth with
$G=0.94\times\frac{2e^{2}}{h}$, indicating that only weak disorder
is present in our samples. At lower temperatures, some small
undulations of the conductance values appear at the plateau, but its
average value is well defined with the standard deviation being much
less than the average value (see error bars in Figs. 3 and 4). A
similar phenomenon, namely the appearance of such structures at
lower temperatures and their disappearance at higher temperatures,
was also recently observed in clean cleaved-edged overgrown wires
\cite{Picciotto}. The variation of the plateau value (approximately
$20\%$) through the wide temperature range ($1\frac{1}{2}$ decades),
allows us to make a meaningful comparison of the data to the
theories derived in the appropriate limit of weak disorder. Fig.
\ref{Inset} shows the measured variation of conductance versus
temperature.

Early theories, particularly those of Kane and Fisher \cite{Kane}
(and of Ogata and Fukuyama \cite{Fukuyama}), proposed that for
relatively small barriers (weak disorder, which is assumed to result
in relatively small corrections), the conductance of a sufficiently
long, single mode 1D spinfull Luttinger liquid system decreases with
temperature in the manner
\begin{equation}
G'(T)=\frac{2e^{2}}{h}\cdot
g\left[1-\left(\frac{T}{T_{0}}\right)^{g-1}\right]. \label{Kane}
\end{equation}
Here, $g<1$ is a dimensionless parameter, which is a measure of the
strength of the interactions. For repulsive interactions, $g$ is
given roughly by the expression
$g=\frac{1}{\sqrt{1+\frac{U}{2E_{F}}}}$, where $U$ is the Coulomb
interaction energy between neighboring electrons. $T_{0}$ is a
parameter describing the strength of the backscattering (disorder)
in the wire; at $T\sim T_{0}$ the corrections to $G(T)$ become of
order $\frac{2e^{2}}{h}$. Both theories predict a correction of
$g\frac{2e^{2}}{h}$ even for ballistic wires at relatively high
temperatures. These imply that for sufficiently long wires, one
cannot observe values close to $\frac{2e^{2}}{h}$ in GaAs, since the
value of $g$ is expected to be of the order of $\simeq0.7$ in such
wires, as was already pointed out by Tarucha \emph{et al.}
\cite{Tarucha}.

This contradiction was also addressed in detail in several
theoretical papers \cite{Maslov1,Oreg,Maslov2,Safi,Oreg2}. According
to the theory of Maslov \cite{Maslov1}, the interaction parameter
$g$ of the wire determines the exponent of the temperature
variation, whereas the pre-factor $g$ in equation (\ref{Kane})
should be set to 1 (noninteracting reservoirs). Fig. \ref{Inset}
(dashed line) shows the curve calculated from this modified
equation.

A different but numerically equivalent result was derived by Oreg
and Finkel'stein \cite{Oreg}. They also demonstrated that for an
infinite clean wire, the conductance keeps the universal value
$\frac{2e^{2}}{h}$ per mode, even in the presence of interactions.
According to their theory, because of the electric field
renormalization by the interactions, the results given by Kane and
Fisher \cite{Kane} of equation (\ref{Kane}) are modified in the
following way:
\begin{equation}
G(T)=\frac{2gG'(T)}{\frac{h}{e^{2}}(g-1)G'(T)+2g}.\label{Oreg}
\end{equation}
As can be easily verified, the leading term in the temperature
variation of the conductance of equation (\ref{Oreg}) leads to the
same results given by Maslov \cite{Maslov1}.

As one can see from Fig. \ref{Inset}, an excellent fit is obtained
for both theories \cite{Maslov1,Oreg}, and we obtain
$g=0.64\pm0.05$, as is expected for electrons in GaAs wires. Indeed,
this value is consistent with the experiments in Ref. \cite{Yacoby1}
, showing $g$ values between $0.66$ and $0.82$. Moreover, using the
Fermi energy $E_{F}\cong1.5\pm0.5~meV$ (half of level spacing
between 1D sub-bands estimated in our previous experiments
\cite{Kaufman1}), we calculate the corresponding electron densities
at the middle of the plateau, obtaining
$n_{_{1D}}=3.2\pm0.5\times10^{5}~cm^{-1}$. Substituting the above
value for $n_{_{1D}}$ into $U=\frac{e^{2}}{\varepsilon}\times
n_{_{1D}}$ we get $U=3.85\pm0.60~meV$ which yields the values of
$g=0.66\pm0.04$, consistent with our fit to the LL model.

\begin{figure}[t]
\centering
\includegraphics[width=1\linewidth]{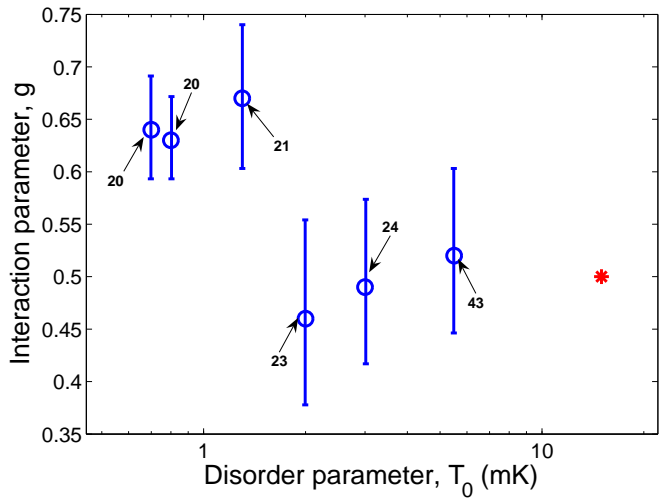}
\caption{Interaction parameter $g$ vs. disorder parameter $T_{0}$.
The values of $\Delta G/G$ (in the temperature range $0.1-4.2~K$)
expressed in \% are shown for each point. The star represents an
estimate for the strength of the disorder of the results reported in
Ref. \cite{Rother}. The wrong values of $g\approx0.5$ (at high
disorder) are established by using the perturbation formula in a
region where it is inapplicable.}\label{Plot}
\end{figure}

Disorder in V-groove QWRs stems mainly from interface roughness
brought about by lithography imperfections on the patterned
substrate and peculiar faceting taking place during MOVPE on a
nonplanar surface \cite{Kapon_a}. The disorder results in potential
fluctuations along the axis of the wire, and manifests itself in
localization of excitons and other charge carriers as evidenced in
optical spectroscopy studies of these wires \cite{Kapon_b}. Optical
and structural studies indicate the formation of localizing
potential wells along the wires with size in the range of several
$10~nm$ \cite{Kapon_c}. The specific features of the disorder in the
QWRs studied here, in terms of depth and size of the localization
potential, are expected to vary from sample to sample.  In fact, the
degree of disorder is represented in our analysis of the temperature
dependence of the conductance by the parameter $T_{0}$. Repeating
the analysis of Fig. \ref{Inset} for several samples, we observed in
all the wires having small amount of disorder, namely showing
$T_{0}<2~mK$, similar values of $g$, namely $g=0.66$. However, other
wires with stronger disorder ($T_{0}>2~mK$), showed lower values of
g, around $g=0.5$. Fig. \ref{Plot} summarizes the values of g vs.
$T_{0}$, obtained for our different wires. The values of the total
change $\frac{\Delta G}{G}$ were calculated for each wire in the
temperature range $0.1-4.2~K$ and are also shown in Fig. \ref{Plot}.
Note that there is an agreement between the two indicators for the
strength of the disorder, $T_{0}$ and $\frac{\Delta G}{G}$. The
transition between $g=0.66$ and $g=0.5$ at $T_{0}\approx2~mK$ occurs
for $\frac{\Delta G}{G}\approx23\%$. We believe that above
$T_{0}\approx2~mK$ the disorder in the wires is strong enough so
that the description by perturbation theory is no longer valid.
Trying to fit such data with perturbation-theory equations gives
inevitably lower (and wrong) values of $g$. For such wires, one
should use other theories, concerning stronger disorder due to many
impurities \cite{Gurnyi}, or stronger backscattering \cite{Fendley}
in the system. The results of conductance measurements in GaAs wires
reported recently by Rother et al. \cite{Rother} also correspond to
highly disordered samples, and also give $g=0.5$. Indeed, analyzing
their data, we estimate the value of $T_{0}\approx 15~mK$ (marked by
a star in Fig. \ref{Plot}) and the change in the conductance
$\frac{\Delta G}{G}\approx 10\%$ over a small temperature range
($1-3~K$). These values are even larger than corresponding values
for our most disordered sample in the same temperature range.

It is highly unlikely that the observed temperature dependence could
be attributed to the contact resistance between the 2DEG and the 1D
subbands outside the gated region for the reasons outlined below:
\newline a) if the contact resistance is treated quantum mechanically \cite{Yacoby2},
namely as a change of the transmission from the 2DEG to the 1D
subbands in the ungated region, we would expect that $\frac{\Delta
G}{G}$ would be similar for any number of 1D channels under the
gate. We however observe that $\frac{\Delta G_{1}}{G_{1}}$ of the
first plateau is much smaller than $\frac{\Delta G_{2}}{G_{2}}$ of
the second plateau at the same temperature range. The latter,
however, is consistent with the expected result of the Luttinger
model when the scattering occurs under the gated region, since the
effect of the Coulomb interaction on the transmission depends on the
number of 1D subbands. Indeed, from an analysis of higher steps in
the conduction depletion curve, used in a smaller temperature range
($0.1-0.6~K$, where the plateaux are better resolved), we deduce the
values $g=0.55$ and $g=0.47$ for the second and the third plateaux,
respectively, which agrees with the theoretical values of $0.54$ and
$0.47$ \cite{Oreg}.
\newline b) if the decrease of the conductance is considered
as an additional contact resistance added in series to the wire
(i.e., treated classically), than the values of the transmission for
each channel at low temperature at the second plateau would increase
with lowering temperature and eventually exceed unity for each
channel. Therefore, we conclude that the observed decrease of the
conductance is due to the interactions in the LL model.

In conclusion, we have measured the temperature dependence of the
electrical conductance in single mode quantum wires. We find that
our data is consistent with theoretical calculations
\cite{Maslov1,Oreg} based on the LL model, in the limit of weak
disorder in the system. We showed that the use of the perturbative
result (namely $G'\propto T^{g-1}$) in order to estimate $g$, is
valid only for wires produced with a \emph{moderate} amount of
disorder ($T_{0}<1~mK$).

We thank Dganit Meidan for constructive discussions of our results.
This research was partially supported by the Israel Science
Foundation, founded by the Israeli academy Sciences and Humanities
Centers of Excellence Program and by ISF grant 845/04.

\end{document}